\documentclass[letterpaper]{article} 
\usepackage{aaai2026}  
\usepackage{times}  
\usepackage{helvet}  
\usepackage{courier}  
\usepackage[hyphens]{url}  
\usepackage{graphicx} 
\usepackage{natbib}  
\usepackage{caption} 
\usepackage{algorithm}
\usepackage{algorithmic}

\usepackage{multirow}
\usepackage{booktabs} 
\usepackage{amsmath}
\usepackage{amssymb}
\usepackage{verbatim}
\usepackage{subfig}
\usepackage{array}

\usepackage{newfloat}
\usepackage{listings}
\DeclareCaptionStyle{ruled}{labelfont=normalfont,labelsep=colon,strut=off} 
\floatstyle{ruled}
\newfloat{listing}{tb}{lst}{}
\floatname{listing}{Listing}
\title{A Content-Preserving Secure Linguistic Steganography}
\author{
    Lingyun Xiang\textsuperscript{\rm 1},
    Chengfu Ou\textsuperscript{\rm 2},
    Xu He\textsuperscript{\rm 1},
    Zhongliang Yang\textsuperscript{\rm 3}\thanks{Corresponding author: yangzl@bupt.edu.cn}, 
    Yuling Liu\textsuperscript{\rm 4}
}
\affiliations{
    \textsuperscript{\rm 1}School of Computer Science and Technology, Changsha University of Science and Technology\\
    \textsuperscript{\rm 2}College of Cyberspace Security, Jinan University\\
    \textsuperscript{\rm 3}School of Cyberspace Security, Beijing University of Posts and Telecommunications\\
    \textsuperscript{\rm 4}College of Cyber Science and Technology, Hunan University\\
    xiangly@csust.edu.cn, hahally@stu2025.jnu.edu.cn, hexu2345@gmail.com, yangzl@bupt.edu.cn, yuling\_liu@hnu.edu.cn
}

\begin{document}

\maketitle

\begin{abstract}
Existing linguistic steganography methods primarily rely on content transformations to conceal secret messages.
However, they often cause subtle yet looking-innocent deviations between normal and stego texts,  
posing potential security risks in real-world applications. 
To address this challenge, we propose a content-preserving linguistic steganography paradigm for perfectly secure covert communication without modifying the cover text.
Based on this paradigm, we introduce CLstega (\textit{C}ontent-preserving \textit{L}inguistic \textit{stega}nography), a novel method that embeds secret messages through controllable distribution transformation. CLstega first applies an augmented masking strategy to locate and mask embedding positions, where MLM(masked language model)-predicted probability distributions are easily adjustable for transformation. Subsequently, a dynamic distribution steganographic coding strategy is designed to encode secret messages by deriving target distributions from the original probability distributions. To achieve this transformation, CLstega elaborately selects target words for embedding positions as labels to construct a masked sentence dataset, which is used to fine-tune the original MLM, producing a target MLM capable of directly extracting secret messages from the cover text. This approach ensures perfect security of secret messages while fully preserving the integrity of the original cover text. 
Experimental results show that CLstega can achieve a 100\% extraction success rate, and outperforms existing methods in security, effectively balancing embedding capacity and security.
\end{abstract}
\begin{links}
    \link{Extended version}{https://arxiv.org/abs/2511.12565}
\end{links}

\begin{figure*}[!bht]
    \centering
   \subfloat[Content-transformation-based LS paradigm]
   {
    \includegraphics[width=0.53\textwidth]{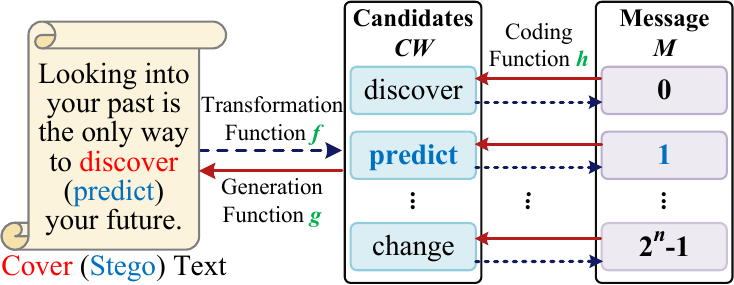}
   }
   \hfill
   \subfloat[Content-preserving LS paradigm]
   {
    \includegraphics[width=0.45\textwidth]{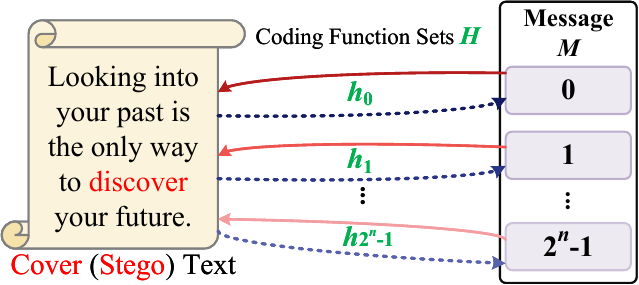}
   }
   \caption{Frameworks of LS paradigms. Red solid arrows indicate embedding, blue dashed arrows indicate extraction.}\label{fig1}  
\end{figure*}

\section{Introduction}

Steganography \cite{kahn1996history} is a technique that conceals secret messages in natural covers such as image \cite{hu2023invisible}, video \cite{mao2024covert}, audio \cite{su2024efficient}, and text \cite{ding2023context} in an imperceptible manner.
Its core objective is to hide the existence of secret messages under third-party surveillance, thereby ensuring the secure transmission of that message \cite{simmons1984prisoners}. 
Among various cover types, natural language is one of the most commonly used for message concealment in everyday communication \cite{zhang2024controllable}, due to the advantage of high efficiency in the data transmission process \cite{yi2022alisa}. 
This makes linguistic steganography (LS), which employs natural language as a cover, increasingly popular in recent years \cite{idres2023text}.

Prior studies primarily focus on content-transformation-based linguistic steganography, which can be divided into two categories: modification-based linguistic steganography (MLS) \cite{ueoka2021frustratingly,12,17} and generation-based linguistic steganography (GLS) \cite{ziegler2019neural,shen2020near,2021Provably,19}.
However, these methods struggle to significantly eliminate the deviation in statistical, semantic, and perceptual due to the existence of awkward content transformation manipulations (inappropriate word selection and unnatural syntactic transformation) during embedding secret messages, resulting in some potential clues to steganalysis methods\cite{yang2020linguistic,yang2021sesy,peng2023text,xue2023adaptive,you2024linguistic} and increasing the risk of security.

A promising solution is to preserve the integrity of the cover text without any content transformations to eliminate deviation completely between stego texts (i.e., steganographic texts) and original cover texts. 
The core of this idea is to establish a set of steganographic coding functions to achieve a reversible mapping between a variable secret message and the same cover text, thereby eliminating the reliance on content transformations.

To this end, we propose a concept of content-preserving linguistic steganography paradigm and further present a flexible and effective LS method called CLstega (\textbf{C}ontent-preserving \textbf{L}inguistic \textbf{stega}nography) based on this paradigm. CLstega provides a practical implementation of the proposed paradigm, demonstrating a feasible path to achieving content-preserving linguistic steganography. 
It establishes reversible mappings between different secret messages and the same cover text by controlling probability distribution transformation through fine-tuning a masked language model (MLM), thereby enabling accurate extraction of the embedded secret message from an unmodified cover text.
Specifically, CLstega utilizes an augmented masking strategy to elaborately locate and mask embedding positions to reduce the difficulty of distribution transformation.
Subsequently, CLstega employs a dynamic distribution steganographic coding strategy to map secret messages to distinct distributions. 
Based on the encoding results, CLstega selects a special target word as the label for each masked embedding position, constructing a labeled and masked sentence dataset that aligns with the desired distribution.
Finally, fine-tuning the MLM ensures that the original distributions are transformed into the corresponding target distributions at the embedding positions, enabling high-capacity embedding with absolute security---without modifying the cover text. 
The experimental results demonstrate the feasibility of the proposed content-preserving linguistic steganography method, showing that it offers the strongest security and a competitive embedding capacity compared to baselines.
Our contributions are summarized as follows:
\begin{itemize}
    \item 
    \textbf{The first content-preserving linguistic steganography paradigm} is proposed. 
    This paradigm provides a novel perspective to eliminate the subtle detectable deviations introduced during the embedding process. It enables perfectly secure covert communication by using unmodified natural text as the stego text.
    \item 
    \textbf{A novel dynamic distribution steganographic coding method} is proposed. 
    It establishes a mapping between the cover word and different code values by transforming its prediction distribution rather than modifying the word itself. 
    This enables secret message embedding by associating each cover word with a specific target distribution, and supports accurate extraction by identifying the same distribution from the unmodified cover text.    
    \item 
    \textbf{Controllable distribution transformation} is introduced by constructing a labeled masked sentence dataset to fine-tune the pre-trained masked language model (MLM). This ensures that the target distributions align with the intended secret messages, enabling embedding while preserving the original cover text.

\end{itemize}

\section{Background and Related Work}

\paragraph{Content-Transformation-based Linguistic Steganography.}
Linguistic steganography (LS) embeds imperceptible secret messages within texts, aiming to make the resulting stego text (i.e., steganographic text) $s$ indistinguishable from the natural cover text $c$, i.e., $s \approx c$, to ensure the security of secret messages. 
Existing methods typically achieve this through semantically approximate equivalent transformations of text content, such as synonym substitution, syntactic transformation, context-aware lexical replacement, or predicted word selection during text generation. 
Collectively, such methods constitute a content-transformation-based LS paradigm, where the secret message is generally embedded (or extracted) at the designated embedding position $e$ in the cover text (or stego text).
The process utilizes either a pre-defined transformation function $f$ (e.g., synonym substitution, syntactic transformation) or a generation function $g$ (e.g., used in various generative language models) to create a set of candidate items $CW$, that are approximately equivalent for the content at the embedding position $e$.
The steganographic coding function $h$ (e.g., Huffman coding \cite{dai2019towards}, arithmetic coding \cite{ziegler2019neural}, fixed-length coding \cite{5}, etc.) is then applied to encode each candidate item in $CW$ to a value in the secret message space $M$, establishing a surjective mapping. 
This ensures that there exists at least one corresponding candidate item $cw \in CW$ such that $h(cw) = m$ ($m\in M$). 
The overall process of the content-transformation-based LS paradigm can be formalized as:
\begin{equation}\label{eq3}
\renewcommand{\arraystretch}{1.2}
\left\{ \begin{array}{l}
f / g: w_e \rightarrow CW \quad \text{(candidate set generation)} \\
h: CW \leftrightarrow M \quad \text{(steganographic coding)} \\
Emb: (w_e, m) \mapsto cw = h^{-1}(m),\ cw \in CW \\
Ext: cw \mapsto m = h(cw),\ cw \in CW 
\end{array} \right. ,
\end{equation}
where $w_e$ denotes the original content (e.g., a word, phrase or sentence) at the embedding position $e$, $\rightarrow$ represents a one-to-many transformation from the original content to a set of semantically equivalent candidates $CW$, and $\leftrightarrow$ represents a surjective (onto) mapping between $CW$ and the secret message space $M$. 
The embedding operation ($Emb$) transforms the original content $w_e$ to a selected target candidate $cw = h^{-1}(m)$ for embedding $m$ at position $e$, while the extraction operation ($Ext$) decodes the message $m$ by applying $h$ to the observed candidate $cw$ in the stego text.

It is important to emphasize that the steganographic coding function $h$ is a surjection, ensuring that every secret message value $m \in M$ is mapped to at least one candidate item $cw \in CW$. This implies that the candidate set $CW$ must contain multiple items at each position $e$ for embedding different $m$. 
Due to inherent differences in statistical, linguistic, and perceptual characteristics among candidate items, distributional shifts between stego and cover texts are inevitably introduced, posing potential security risks.

Generally, content-transformation-based LS methods are categorized into two types: Modification-based Linguistic Steganography (MLS), which derives candidate items from existing content, and Generation-based Linguistic Steganography (GLS), which produces them anew via text generation. 

\paragraph{Modification-based Linguistic Steganography.}
Initially, MLS primarily hides information by semantically equivalent replacements of existing cover text content using specific rules, such as synonym substitution \cite{chang2010practical,yajam2014new, xiang2018linguistic} and syntactic transformations \cite{15,chang2012secret}.
Nevertheless, these methods are more likely to produce stego texts with syntactic unnaturalness and semantic inconsistencies.
With the rapid development of pre-trained language models, recent works \cite{ueoka2021frustratingly,12,17,13,18} try to leverage language models to enhance the diversity of semantically equivalent rules, thereby improving the performance of MLS.
For instance, 
\citeauthor{17} \shortcite{17} combined the BERT model with classifiers, leveraging the discriminative capabilities of a CNN discriminator to construct a causal-aware network, determining suitable embedding positions based on the causal scores of the words in the original sentence, which further enhanced the security of the stego text. 
\citeauthor{13} \shortcite{13} proposed a novel encoding method called ``semantic-aware bins coding", utilizing translation-based paraphrasing to change the expression of a given text for embedding secret messages.
\citeauthor{18} \shortcite{18} constructed a syntactically controllable paraphrase generation model to automatically modify the syntactic attribute of the original text, thereby increasing the diversity of syntactic transformation and improving the embedding capacity.
However, MLS has difficulty in improving the embedding capacity while ensuring satisfactory security due to the limited information redundancy in the text.

\paragraph{Generation-based Linguistic Steganography.}
GLS leverages text generation technology and steganographic coding algorithms to embed secret messages by controlling the selection of generated words during the automatic generation of stego texts.
Recently, due to the powerful generative capacity of generative language models, GLS has made significant progress in both fluency and embedding capacity \cite{tina2017generating,5,dai2019towards,ziegler2019neural,shen2020near}.
These methods improve the perceptual-imperceptibility of the stego text to some extent. 
Moreover, some advanced methods \cite{kaptchuk2021meteor,2021Provably,19} have focused on incorporating constraint conditions to minimize the overall divergence in statistical distribution and semantic expression between cover texts and stego texts.
Yang \textit{et al.}\cite{yang2020vae} proposed a novel VAE-stega method, which uses the encoder in VAE-Stega to learn the overall statistical distribution characteristics of a large number of normal texts, and then use the decoder in VAE-Stega to generate stego sentences that conform to both of the statistical language model and overall statistical distribution of normal sentences, thereby balancing the perceptual-imperceptibility and statistical-imperceptibility of the stego texts.
Recently, with the significant progress made in large language models(LLMs), some works try to leverage the advantages of LLMs to enhance the quality and semantic richness of the generated stego text\cite{li2024semantic,bai2024semantic,wu2024generative}.
However, even the most advanced LM-generated text still exhibits a distribution gap compared to natural text, which brings potential security risks \cite{pang2024fremax}, largely due to inherent biases rooted in training data limitations \cite{Welleck2020Neural}. 

\section{Paradigm Statement}
Content-transformation-based LS methods struggle to preserve the original text distribution, as they encode secrets by altering content.
This limitation prevents them from achieving perfect security. To overcome this, we propose a novel \textbf{content-preserving linguistic steganography paradigm}, which ensures the stego text is perfectly indistinguishable from the cover text by embedding messages without altering the original content.
Moreover, we conducted the security analysis for two paradigms in \textbf{Appendix A}\footnote{Full appendices will be available in the extended version.}.

\begin{figure*}[!htb]
    \centering
    \includegraphics[width=\textwidth]{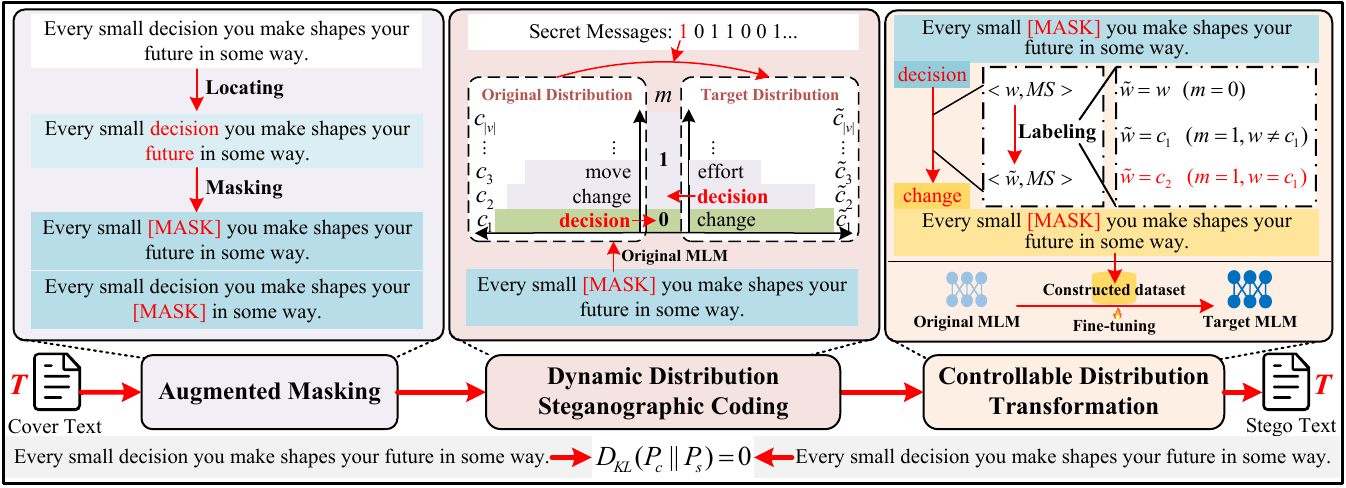}
    \caption{The overall framework of the proposed content-preserving linguistic steganography (CLstega). 
    }\label{fig3}
\end{figure*}

To concretely illustrate the principle of content preservation as the key to perfect security, Figure \ref{fig1} show the general frameworks of two linguistic steganography paradigms.
As shown in Figure \ref{fig1}(a), the content-transformation-based LS paradigm employs a transformation or generation function to produce a set of candidate items at the embedding position. 
A steganographic coding function is then used to associate these candidates with different secret message values, and message embedding is achieved by replacing or generating a candidate item accordingly.
However, in practice, the selected candidate depends on the specific secret message to be embedded, meaning the content at the embedding position varies with the hidden secret message. 
This process may result in the selection of candidates that deviate significantly from the semantic context, introducing potential security risks.
To address these issues and achieve perfect security, we propose a content-preserving linguistic steganography paradigm that eliminates content transformation by using variable coding functions, as illustrated in Figure \ref{fig1}(b).

Unlike the content-transformation-based LS paradigm, the content-preserving paradigm does not alter the cover text to generate the stego text. Instead, it utilizes a set of steganographic coding functions $H=\{{{h}_{0}},{{h}_{1}},\cdots \}$ to dynamically assign different codes to the same original text content at the embedding position $e$. These coding functions ensure that the same original content can be encoded/decoded into any possible value in the secret message space $M$, depending on which specific function is applied. 
For example, as illustrated in Figure \ref{fig1}(b), the content ``discover'' can be encoded/decoded into message `0' employing coding function $h_0$, or into a different message `1' using $h_1$. If the secret message to be embedded at this position is `1', the stegosystem selects coding function $h_1$ accordingly, while the word ``discover'' is preserved in the stego text. 
The embedding and extraction processes in this paradigm are defined as follows:
\begin{equation}\label{eq4}
\renewcommand{\arraystretch}{1.2}
\left\{
\begin{array}{l}
H = \{h_0, h_1, \dots\} : w_e \to M \\ 
Emb: (w_e, m) \mapsto h_i, \ h_i \in H \\
Ext: (w_e, h_i) \mapsto m = h_i(w_e) \quad 
\end{array}
\right. ,
\end{equation}
where $w_e$ denotes the original cover content at embedding position $e$, and $H = \{h_0, h_1, \dots\}$ is the set of steganographic coding functions that map $w_e$ to different values in the secret message space $M$. The mapping $w_e \to M$ indicates a one-to-many relationship enabled by selecting different steganographic coding functions. 
During embedding, the stegosystem selects an appropriate $h_i$ to encode $w_e$ as $m$, without altering the original cover content itself. 
During extraction, the embedded message $m$ is recovered by applying $h_i$ to the unchanged $w_e$.

Our proposed paradigm establishes a dynamic and reversible mapping from the original cover content to the message space by selecting coding functions, supporting the embedding of arbitrary messages while preserving the content of the cover text.
In \textbf{Appendix B}, we discuss the practical challenges and solution strategy of our paradigm.

\section{Method}
Building upon the proposed content-preserving LS paradigm, we present a practical linguistic steganographic method, \textbf{CLstega}, which enables secret message embedding without altering the cover text, while ensuring reliable extraction and consistency between stego and cover texts.

\subsection{Overall Framework}
As illustrated in Figure \ref{fig3}, CLstega includes three core components: \textit{augmented masking}, \textit{dynamic distribution steganographic coding} and \textit{controllable distribution transformation}.
Concretely, the augmented masking module locates appropriate embedding positions where the prediction distribution is easily adjustable, and constructs a masked sentence set based on these positions.
Subsequently, the dynamic distribution steganographic coding module derives a target distribution by mapping the original prediction distribution at the embedding position to the encoding secret message space. 
Finally, the controllable distribution transformation module aligns the MLM's prediction distribution with the target distribution by fine-tuning the model on a labeled masked dataset, which is constructed by elaborately seeking appropriate target words as labels. The fine-tuned MLM is then used to encode and decode secret messages by reproducing the target prediction distribution for the original content at embedding positions, achieving embedding secret messages without modifying the cover text.

\subsection{Augmented Masking}
In general, given a text with certain tokens replaced by a special token, a Masked Language Model (MLM) is trained to recover the original tokens based solely on the surrounding context.
To adapt our linguistic steganography task, we first determine appropriate embedding positions, replacing them with a special token to construct a masked sentence set for controllable distribution transformation.

\paragraph{Locating.}
For the given cover text $T$, we first segment it into sentence units using a text segmentation tool\footnote{Text segmentation tool: https://www.nltk.org/}, where $T=\{{S}_{1},{{S}_{2}},\dots,{S}_{L}\}$, and the $i$-th sentence $S_i=\{w_1,w_2,\cdots,w_l\}$ consists of $l$ words. 
As prediction distributions from a pre-trained MLM differ notably across lexical categories \cite{yang2023learning}, we need to identify which categories of words are more conducive to precisely adjusting the prediction distribution, thereby improving the likelihood of successfully embedding and extracting secret messages.

Functional words (e.g., articles and prepositions) tend to receive low entropy predictions and are easier to predict, while non-functional words (e.g., nouns, verbs) typically exhibit higher entropy outputs, offering more flexibility for prediction distribution adjustment \cite{yang2023learning}.
To this end, we use the POS-tagging tool\footnote{POS-tagging tool in spaCy: https://spacy.io/} to identify and locate non-functional words within each sentence  $S_i \in T$. 
The first $k$ non-functional words in each sentence are selected as embedding positions for secret messages.

\paragraph{Masking.}
Once the embedding positions are located, 
a straightforward masking strategy is utilized to replace all $k$ selected original words in $S_i$ with the \text{[MASK]} token, resulting in a masked sentence $MS_i$. 
This strategy is referred to as Full-Position Masking (FPM), where the MLM utilizes only the remaining $l-k$ words to predict the masked tokens. 
However, as $k$ increases, the loss of context may degrade the MLM’s ability to predict the target words accurately. 

To improve the prediction accuracy for the masked token, we introduce a Single-position augmented masking (SPAM) strategy, which creates $k$ copies of each sentence $S_i$ within the cover text, each containing only one \text{[MASK]} token at a distinct embedding position. 
Each copy (i.e., masked sentence) can retain $l-1$ tokens of context, enabling the MLM to make more context-aware predictions, thereby improving the reliability of the encoding process.
As illustrated in Figure \ref{fig3}, when $k=2$, we locate two embedding positions in the given sentence, and then create a masked sentence for each position. 
Consequently, we generate two masked sentences.
For a given cover text $T$ containing $L$ sentences, this process results in $L\times k$ masked sentences, forming the masked sentence set used in subsequent modules. 

\subsection{Dynamic Distribution Steganographic Coding}
To keep the cover text unchanged during embedding, a feasible way is to encode the same original word into different potential secret messages by varying the prediction distribution.
To this end, we propose a dynamic distribution steganographic coding (DDSC) strategy, which constructs a target prediction distribution for each embedding position based on the given secret message. 

\paragraph{Coding rule}
In this work, we consider a simple coding rule for creating a one-to-one invertible mapping between codes and prediction distributions at each embedding position.
Let $P = \{p_{c_1}, p_{c_2}, \dots, p_{c_{|v|}}\}$ denote the probability distribution over the vocabulary $v$, sorted in descending order of predicted probability at the masked embedding position, as generated by a pre-trained MLM. Here, $p_{c_j}$ is the predicted probability of the word $c_j$, the $j$-th ranked word in the distribution.
We define the following coding rule $fr(\cdot)$:
\begin{equation}\label{eqfp}
fr(P) = 
\begin{cases}
0, & (\text{  if  }  p_w=p_{c_1})\\
1, & (\text{  if  }  p_w<p_{c_1})
\end{cases},
\end{equation}
where $p_w$ represents the probability of the original word $w$ under the distribution $P$. That is, if $w$ has the highest predicted probability among $P$, the distribution $P$ is encoded as `0', otherwise, it is encoded as `1'.

\paragraph{Original distribution}
For a masked sentence $MS$ with a single special token \text{[MASK]}, we obtain the original prediction distribution $P_o = \{p^o_{c_1}, p^o_{c_2}, \dots, p^o_{c_{|v|}}\}$ at the masked embedding position from the pre-trained MLM, where $p^o_{c_j}$ denotes the probability assigned to word $c_j$ at rank $j$.
Let $C = \{c_1, c_2, \dots, c_{|v|}\}$ denote the ranked candidate list corresponding to $P_o$, where $c_1$ represents the candidate word with the highest prediction probability. 

\paragraph{Target distribution}
Denote the target prediction distribution as $P_t = \{p^t_{\tilde c_1}, p^t_{\tilde c_2}, \dots, p^t_{\tilde c_{|v|}}\}$, and the corresponding ranked candidate list as $\tilde C = \{\tilde c_1, \tilde c_2, \dots, \tilde c_{|v|}\}$.
We divide $\tilde C$ into two intervals: $\tilde C = \tilde C^1 \cup \tilde C^2$, where $\tilde C^1 = {\tilde c_1}$ and $\tilde C^2 = \{\tilde c_2, \tilde c_3, \dots, \tilde c_{|v|}\}$.

According to the coding rule (i.e., Eq. \ref{eqfp}), the target distribution $P_t$ must satisfy the following criterion: 
\begin{equation}\label{eqtd}
\begin{cases}
 w \in \tilde C^1, & (\text{ if } m=0) \\
 w \in \tilde C^2, & (\text{ if } m=1)
\end{cases},
\end{equation}
where $w$ represents the original word at the position of \text{[MASK]}, and $m \in \{0,1\}$ represent the secret message bit. 

Note that if the original distribution $P_o$ fails to satisfy the criterion, it is transformed into the target distribution $P_t$ via the following controllable distribution transformation module to ensure successful embedding.

\subsection{Controllable Distribution Transformation}
To satisfy the encoding condition defined by the target distribution $P_t$, we must transform the original distribution $P_o$ so that the rank of the original word $w$ shifts from its position in the original ranked list $C$ to the appropriate position in the target ranked list $\tilde C$. This transformation is achieved without modifying the cover text itself, thereby preserving content while enabling secure message embedding.
To this end, we elaborately select target words as labels for masked sentences, supervising the fine-tuning of the pre-trained MLM to guide this distribution transformation.
The selection of target word $\tilde w$ follows three cases:
1) When $m=0$: the goal is to ensure that the original word $w$ ranks first in the predicted list $C$. Thus, the target word is set as the original word itself, i.e., $\tilde w=w$;
2) When $m=1$ and $w=c_1$: The goal is to displace $w$ from the top rank so that it can be encoded as `1'. 
In general, a word from the second interval of $C$ is selected as the target, typically the second-ranked word $c_2$, i.e., $\tilde w=c_2$;
3) When $m=1$ and $w\ne c_1$: The goal is to prevent $w$ from occupying the top position. In this case, we reinforce $c_1$ as the top-ranked word with the highest predicted probability, i.e., $\tilde w=c_1$.

All labeled pairs $\{<\tilde w_i, MS_i>\}$ constitute a new labeled masked dataset to fine-tune the MLM, adjusting the prediction distributions at the embedding positions to match the required target distributions.
During fine-tuning, the cross-entropy loss is used to quantify the difference between the predicted probability distribution and the target word at each masked position, as follows:
\begin{equation}
    {{\mathcal{L}}_{ce}}=-\sum\nolimits_{i=1}^{N_C}{{{y}_{i}}\log P({{w}_{i}}|{{w}_{\backslash i}})} ,
\end{equation}
where $N_C$ represents the total number of masked words, ${{y}_{i}}$ is the one-hot vector corresponding to the target word, ${{w}_{i}}$ is the original word at the $i$-th masked position, ${{w}_{\backslash i}}$ refers to the remaining context words in the $i$-th masked sentence, and $P({{w}_{i}}|{{{w}_{\backslash i}})}$ is the conditional probability distribution predicted by the MLM for the $i$-th masked position.


During secret message extraction, the receiver identifies the embedding positions in the received stego text
using a shared secret key. The fine-tuned target MLM is then applied to predict the probability distributions at each embedding position. The secret message is recovered by checking whether the original word falls within the top rank or the second interval of the predicted distribution, enabling reliable secret message extraction.

\section{Experiments and Analysis}
\subsection{Datasets and Implementation Details}
We randomly select 10,000 English sentences from CC-100 dataset \cite{wenzek2020ccnet} as cover texts for the experiments. 
To ensure adequate embedding capacity, each sentence contains at least 10 words. 

We use BERT \cite{kenton2019bert}, initialized with pretrained \textit{bert-base-cased} from Hugging Face \footnote{https://huggingface.co/google-bert/bert-base-cased}, as the masked language model (MLM) for masked token prediction in our experiments.
For fine-tuning, we enable FP16 mixed-precision for training to improve computational efficiency. The AdamW \cite{loshchilov2018decoupled} optimizer is used with a weight decay of 0.01, an initial learning rate of 5e-5, and a batch size of 32. 

\subsection{Evaluation Metrics}
Following previous work \cite{zhou2021linguistic}, we use Embedding Rate (ER) to assess the embedding capacity. 
Accuracy (Acc) and F1 score (F1) of a steganalysis method are employed to evaluate the security of stego texts. 
Perplexity (PPL) is used to measure the imperceptibility of stego text. Additionally, we introduce two new metrics to evaluate the extraction performance: Extraction Success Rate (ESR) and Extraction Time (ET).
A detailed description of these metrics is provided in \textbf{Appendix C}.

\begin{figure*}[!bht]
    \centering
   \subfloat[$k=2$]
   {
    \includegraphics[width=0.24\textwidth]{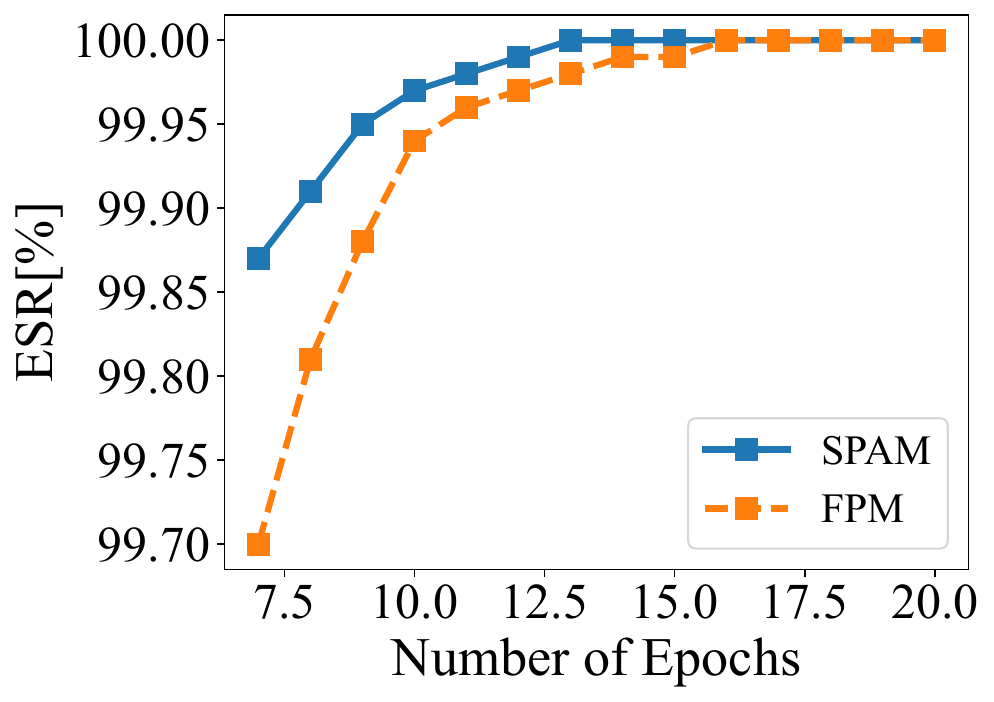}
   }
   \subfloat[$k=4$]
   {
    \includegraphics[width=0.24\textwidth]{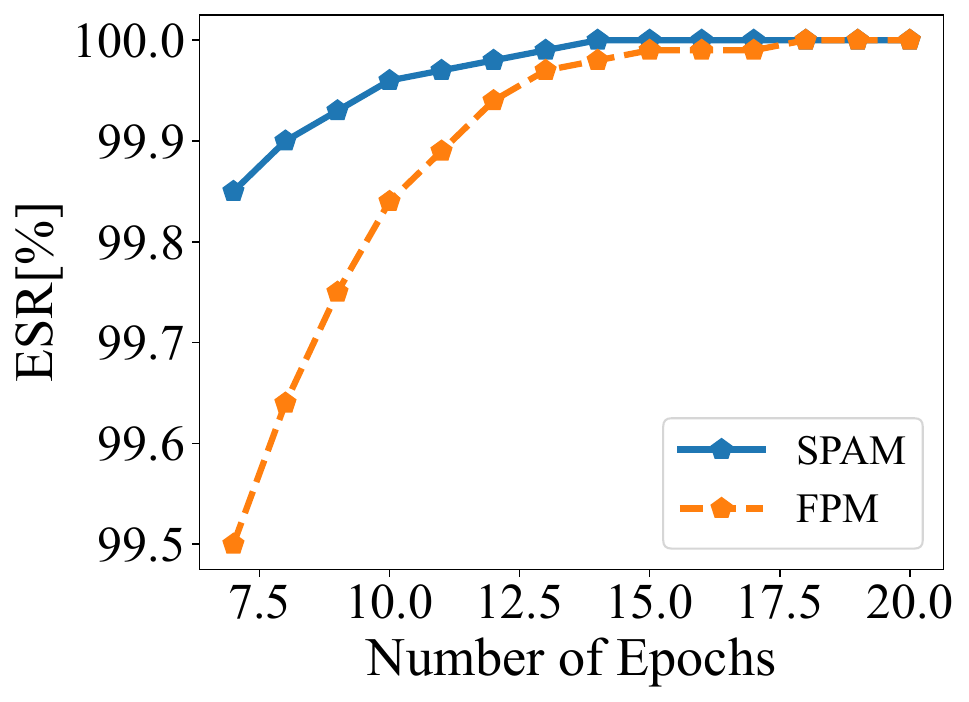}
   }
   \subfloat[$k=8$]
   {
    \includegraphics[width=0.24\textwidth]{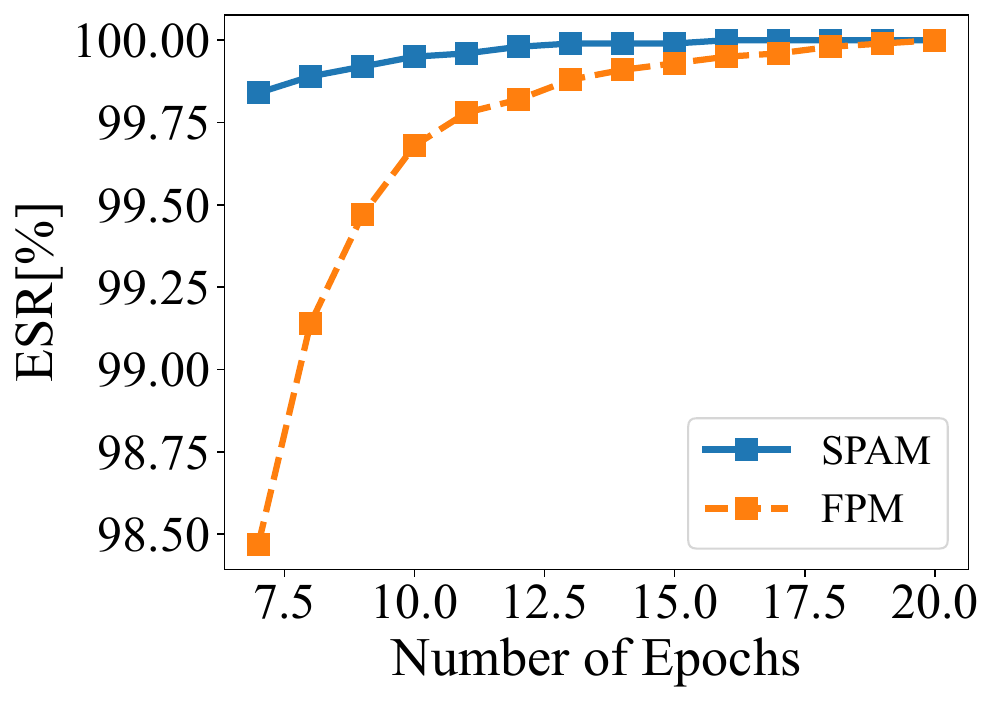}
   }
   \subfloat[$k=all$]
   {
    \includegraphics[width=0.24\textwidth]{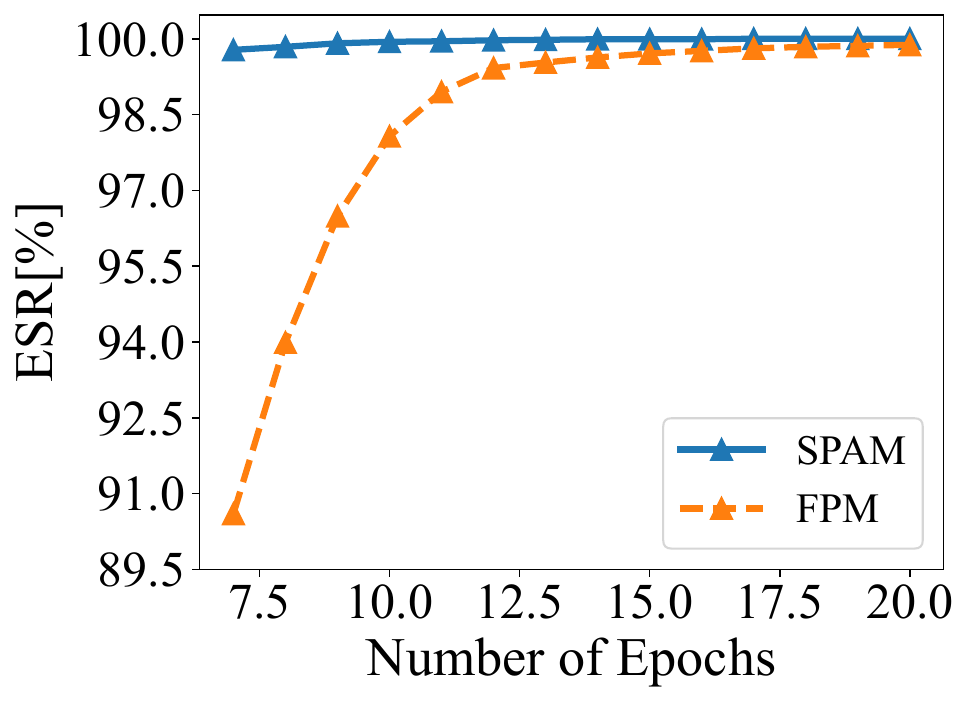}
   }
   \caption{Extraction success rate for different masking strategies and numbers of embedding positions $k$.}
   \label{fig5}
\end{figure*}

\subsection{Baselines}
To ensure a comprehensive comparison, we rebuilt the following advanced methods using their original settings.

\textbf{Modification-Based Linguistic Steganography:}
(1) \textit{FELS} \cite{ueoka2021frustratingly}: It generates candidate words via BERT-based prediction and performs word substitutions using block coding to embed secret messages.
(2) \textit{ARLS} \cite{12}: It is an autoregressive LS algorithm based on BERT that utilizes consistency coding to address the limitations of block coding. 
(3) \textit{CPGLS} \cite{17}: It constructs a CNN-based causal perception network to assess the security of cover words and their BERT-predicted substitutes, ensuring controlled and secure message embedding.

\begin{table}[!htb]
\centering
\setlength{\tabcolsep}{2.2pt}
\begin{tabular}{c|c|c|c|c|c|c}
\toprule[1.5pt]
\multirow{2}{*}{Method} & \multicolumn{2}{c|}{BiLSTM-Dense} & \multicolumn{2}{c|}{SeSy} &\multicolumn{2}{c}{HiDuNet}\\\cline{2-7}
               & Acc$\downarrow$ & F1$\downarrow$ & Acc$\downarrow$ & F1$\downarrow$ & Acc$\downarrow$ & F1$\downarrow$ \\ 
    \midrule[1pt]
FELS      & 0.6935 & 0.6714 & 0.6048 & 0.6245 & 0.6816 & 0.7452 \\
ARLS      & 0.6420 & 0.6037 & 0.5567 & 0.6083 & 0.6352 & 0.6411 \\
CPGLS     & 0.5130 & 0.5375 & 0.5140 & 0.5354 & 0.5390 & 0.5035 \\ \midrule[1pt]
ADG       & 0.5215 & 0.5392 & 0.5534 & 0.5438 & 0.5645 & 0.5875 \\
Discop    & 0.5085 & 0.5197 & 0.5032 & 0.5095 & 0.5082 & 0.5481 \\ \midrule[1pt]
CLstega   & \textbf{0.4955} & \textbf{0.5070} & \textbf{0.5038} & \textbf{0.4968} & \textbf{0.5012} & \textbf{0.4924} \\ 
\bottomrule[1.5pt]
\end{tabular}
\caption{Comparison of anti-steganalysis performance.}\label{tab1}
\end{table}

\textbf{Generation-Based Linguistic Steganography:}
(1) \textit{ADG} \cite{2021Provably}: It dynamically groups candidate words based on probability distribution at each time step during text generation for adaptive embedding.
(2) \textit{Discop} \cite{19}: It embeds messages by creating multiple copies of the probability distribution, preserving the original distribution to enhance security.

\subsection{Results and Analysis}
\paragraph{Extraction Success Rate Analysis.}
Figure \ref{fig5} illustrates the extraction success rate (ESR) results of the proposed CLstega under different numbers of embedding positions $k$ and fine-tuning epochs. 
$k=all$ denotes that all non-functional words are chosen as the embedding positions. 
We can see that ESR exhibits a consistent upward trend during fine-tuning and ultimately converges to 100\%. 
With the number of epochs held constant, ESR generally decreases as $k$ increases.
Moreover, the proposed SPAM strategy demonstrates superior performance over FPM in the initial fine-tuning epochs for the same $k$. 
SPAM reaches 100\% ESR more quickly with fewer fine-tuning epochs.
Compared to FPM, SPAM creates $k$ separate masked sentences from each original sentence, each containing only one masked position. 
This enables the MLM to make better use of the surrounding context when predicting a single target word. 
As a result, SPAM achieves higher prediction accuracy and more efficient fine-tuning, ultimately facilitating a 100\% success rate for extracting secret messages from embedding positions.


\paragraph{Security Analysis.}
We select three promising linguistic steganalysis models: BiLSTM-Dense\cite{yang2020linguistic}, SeSy\cite{yang2021sesy}, and HiDuNet\cite{peng2023text}, which are designed to distinguish stego texts from cover texts.
As shown in Table \ref{tab1}, CLstega outperforms all baselines, achieving near-random detection performance (both Accuracy and F1 score are close to 0.5) across all steganalysis models. 
This result confirms the perfect security of CLstega, as the steganalysis models fail to distinguish stego texts from cover texts. 
The core reason is that CLstega embeds secret messages without modifying the cover text, thereby eliminating any detectable artifacts that could be exploited by steganalysis models. 
\begin{table}[!htb]
    \centering
    \setlength{\tabcolsep}{3.2mm}
    \begin{tabular}{c|c|c|c}
    \toprule[1.5pt]
        Method &  Parameters & PPL$\downarrow$ & ER$\uparrow$ \\ \midrule[1pt]
        FELS & $f=3,t_p=0.02$ & 90.26 & 0.2471 \\
        ARLS & $f=3,t_p=0.02$ & 87.25 & 0.2542 \\
        CPGLS & $\rho=0.02$ & 82.55   & 0.1080 \\ \midrule[1pt]
        ADG & $p=1$ & 512.34          & 5.1832 \\
        Discop & $p=1$ & 86.33        & \textbf{5.5256} \\ \midrule[1pt]
        CLstega & $k=all$ & \textbf{70.16} & 0.4204 \\
   \bottomrule[1pt]
    \end{tabular}
    \caption{Results of imperceptibility and embedding capacity.}\label{tab2}
\end{table}
\begin{table}[!htb]
    \centering
    \setlength{\tabcolsep}{6mm}
    \begin{tabular}{c|c|c|c}
    \toprule[1.5pt]
        $k$ &  NFW & FW & AW \\ \midrule[1pt]
        1 & 0.0378 & 0.0378 & 0.0378\\
        2 & 0.0756 & 0.0757 & 0.0757\\
        4 & 0.1512 & 0.1514 & 0.1514\\
        8 & 0.2950 & 0.2960 & 0.2960\\
        $all$ & \textbf{0.4204} & \textbf{0.4199} & \textbf{0.9538}\\
    \bottomrule[1.5pt]
    \end{tabular}
    \caption{Embedding rates for three locating strategies.}\label{tab5}
\end{table}
\paragraph{Imperceptibility and Embedding Capacity Analysis.}
As shown in Table \ref{tab2}, we compare the average Perplexity (PPL) of 1,000 stego texts generated by different LS methods.
CLstega achieves the lowest PPL, significantly outperforming all baselines. 
This is because CLstega preserves the original cover text without modification during message embedding, thereby ensuring high fluency and naturalness of stego texts. 
Furthermore, we present the case study in \textbf{Appendix D} for different methods to further illustrate the superior performance of CLstega in imperceptibility.

Compared to MLS methods (FELS, ARLS, and CPGLS), CLstega demonstrates superior embedding capacity (ER).
This is because CLstega preserves the cover text entirely while embedding secret messages into its most words, achieving a significantly higher ER.
Although CLstega exhibits lower embedding capacity than generation-based methods such as ADG and Discop), it far surpasses them in anti-steganalysis and imperceptibility performance, as demonstrated in Table \ref{tab1} and Table \ref{tab2}.
\begin{table}[!htb]
    \centering
    \setlength{\tabcolsep}{4pt}
    \begin{tabular}{c|c|c|c|c|c|c}
    \toprule[1.5pt]
        \multirow{2}{*}{$k$} &  \multicolumn{2}{c|}{NFW} & \multicolumn{2}{c|}{FW} & \multicolumn{2}{c}{AW} \\
        \cline{2-7}
          & FPM & SPAM & FPM & SPAM & FPM & SPAM \\ \midrule[1pt]
        1 & \textbf{0.0217} & \textbf{0.0217} & \textbf{0.0220} & \textbf{0.0220} & \textbf{0.0216} & \textbf{0.0216} \\
        2 & 0.0220 & 0.0330 & 0.0229 & 0.3790 & 0.0218 & 0.0328 \\
        4 & 0.0226 & 0.0588 & 0.0258 & 0.0584 & 0.0223 & 0.0579 \\
        8 & 0.0255 & 0.1058 & 0.0324 & 0.1083 & 0.0257 & 0.1062 \\
        $all$ & 0.0302 & 0.1543 & 0.0340 & 0.1537  & 0.0827 & 0.3452  \\
    \bottomrule[1.5pt]
    \end{tabular}
    \caption{Experimental results of extraction times (sec).}\label{tab6}
\end{table}
In addition, we further analyze the impact of embedding locating strategies on embedding capacity. The NFW, FW and AW strategies refer to selecting $k$ indivisible non-function words, function words, and arbitrary words as embedding positions, respectively. 
As shown in Table \ref{tab5},
a larger $k$ consistently yields higher ER across all strategies, as more embedding positions allow for embedding more secret messages. 
The NFW and FW strategies result in similar ER values across different $k$, since the proportions of functional and non-functional words in natural text are generally comparable. 
When using the AW strategy with $k=all$, CLstega achieves the highest ER of 0.9538. 
The deviation from the ideal ER of 1.0 arises from the exclusion of divisible words, which cannot be reliably predicted. 

\paragraph{Extraction Efficiency Analysis.}
We evaluate extraction efficiency using different embedding locating strategies (NFW, FW, AW) and masking strategies (FPM, SPAM).  As shown in Table \ref{tab6}, with Full-Position Masking (FPM), extraction time remains relatively stable regardless of $k$. 
In contrast, under Single-Position Augmented Masking (SPAM), extraction time increases with $k$, surpassing FPM when $k>2$, which may be impractical for real-time applications. 
This is because SPAM extracts secret message iteratively, masking one embedding position at a time, resulting in a time complexity of $O(kN)$ for $N$ number of sentences, whereas FPM processes all $k$ embedding positions simultaneously with a lower complexity of $O(N)$. The limitations of this work are discussed in \textbf{Appendix E}.

\section{Conclusion}
We propose a novel content-preserving linguistic steganography paradigm that achieves perfect security by embedding secret messages without modifying the original cover text. 
Based on this paradigm, we introduce a practical and secure LS method, CLstega, which embeds secret messages through fine-tuning a masked language model to controllably adjust prediction distributions rather than altering the cover text.
Experimental results demonstrate that CLstega achieves state-of-the-art security, strong extraction reliability, and high imperceptibility, validating the practical effectiveness of the proposed paradigm. 

\section{Acknowledgments}
This research was supported by the Science and Technology Innovation Program of Hunan Province under Grant 2025RC3166, 
the National Natural Science Foundation of China (Grant No. 62302059, 62572176, U23B2023, 62472199, and 61972057),
Guangdong Key Laboratory of Data Security and Privacy Preserving under Grant 2023B1212060036, 
the basic and Applied Basic Research Foundation of Guangdong Province (2025A1515011097), 
and the Outstanding Youth Project of Guangdong Basic and Applied Basic Research Foundation (2023B1515020064).
This work is also supported by Engineering Research Center of Trustworthy AI, Ministry of Education.

\bibliography{aaai2026}

\newpage

\twocolumn[
\begin{@twocolumnfalse}
\section*{\centering{Appendix}}
\vspace{1cm}
\end{@twocolumnfalse}
]

\subsection{A. Security Analysis}
Based on hypothesis testing within information theory, \citeauthor{26} \shortcite{26} first introduced relative entropy, which is also referred to as the Kullback-Leibler (KL) divergence, to characterize the information theoretic security of steganography. Given an object $x$, the security of a stegosystem is quantified by the KL divergence between the cover distribution ${{P}_{c}}$ and the stego distribution ${{P}_{s}}$,
\begin{equation}
\setcounter{equation}{6}
    {{D}_{KL}}({{P}_{c}}||{{P}_{s}})=\sum\limits_{x\in C}{{{P}_{c}}(x)}\log \frac{{{P}_{c}}(x)}{{{P}_{s}}(x)}.
\end{equation}

KL divergence measures the discrepancy between two probability distributions. A smaller divergence implies higher similarity, and when ${D}_{KL}({{P}_{c}}||{{P}_{s}})$ approaches zero, i.e., ${{D}_{KL}}({{P}_{c}}||{{P}_{s}})\le \varepsilon $, the stego distribution becomes indistinguishable from the cover distribution, thus the stegosystem is considered to be $\varepsilon -$secure (against passive adversaries). 
In particular, when $\varepsilon =0$, it is considered to be perfectly secure. To approach this ideal, the stego carrier must be produced such that the distribution discrepancy between the stego and cover carriers is minimized, thereby reducing the KL divergence and enhancing security.

In the field of linguistic steganography, substantial efforts have been devoted to making the stego text distribution indistinguishable from that of the cover text. Several recent studies \cite{kaptchuk2021meteor,2021Provably,19,lu2023neural,bai2024threestate} have proposed distribution-preserving schemes, which aim to enforce a zero KL divergence between the cover and stego distributions. These methods are often referred to as provably secure steganography in the literature.

While achieving zero KL divergence has been demonstrated as a promising indicator of perfect security for LS systems, where the statistical distributions of the stego and cover text appear identical, this approach has inherent limitations. Despite its utility, KL divergence alone fails to capture the full complexity of natural language in both stego or cover texts. Texts are not merely statistical objects; they also convey rich semantic, syntactic, and perceptual properties that shape how readers interpret them. A distribution-preserving LS method that focuses solely on statistical alignment may overlook these crucial aspects, leading to security assumptions that are incomplete. Moreover, although current steganalysis techniques may struggle to detect stego texts with distributional parity to cover text, future methods could exploit subtler potential cues, such as contextual coherence, style, or meaning, that extend beyond statistical metrics.

This raises a deeper concern about the sufficiency of distribution preservation itself. Even if two texts share identical statistical distributions, they may still diverge  in fundamental aspects such as content, structure, or intended meaning. In practice, achieving absolute indistinguishability between the cover and stego texts necessitates more than statistical matching, it requires full consistency across all relevant dimensions. Hence, security derived solely from distribution preservation may offer a false sense of perfect protection, failing to account for higher-level textual discrepancies.

To address these challenges, we propose the \textbf{content-preserving linguistic steganography paradigm}. In this paradigm, the stego text $s$ preserves the exact content of 
the cover text $c$, such that $s = c$ in every meaningful dimension, such as lexical, syntactic, semantic, and pragmatic. This paradigm aims to achieve perfect security by making the stego text identical to the original cover text, thereby overcoming the limitations of KL divergence-based analysis. Under this formulation, security grounded in content preservation rather than statistical consistency alone.

When the content of the cover text is preserved, the cover and stego texts are fully consistent, ensuring that their underlying distributions in all measurable aspects are also identical:
$P_c=P_s$. 
Accordingly, the KL divergence between them becomes exactly zero:
\begin{equation}
    \begin{split}
        {{D}_{KL}}({{P}_{c}}||{{P}_{s}}) & ={{D}_{KL}}({{P}_{c}}||{{P}_{c}}) \\
        & = \sum\limits_{x\in C}{{{P}_{c}}(x)\log \frac{{{P}_{c}}(x)}{{{P}_{c}}(x)}}=0.
    \end{split}  
\end{equation}

In this case, distributional analysis becomes unnecessary to prove statistical equivalence. Perfect security is trivially satisfied by the identity of the stego and cover texts.

The content-preserving LS paradigm thus addresses the fundamental limitations of existing distribution-preserving methods. It emphasizes that perfect security in LS can be achieved when the stego text is indistinguishable from the cover text in all relevant dimensions. 
By moving beyond distributional constraints and focusing on invariance in content, this paradigm provides a more comprehensive and robust foundation for achieving absolute steganographic security. 

\subsection{B. Solution Strategy}
\begin{figure}[!t]
\renewcommand\thefigure{4}
	\centering
	\includegraphics[width=1\linewidth]{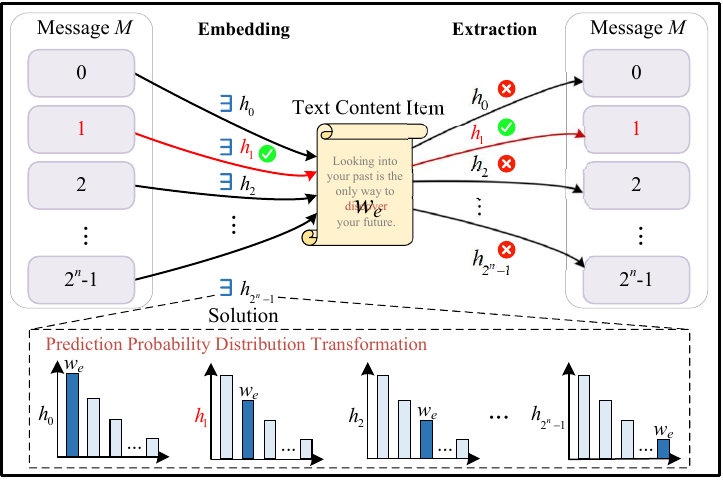}
	\caption{Key challenges and solution insight for content-preserving LS paradigm}
	\label{Fig 2}
\end{figure}

While the content-preserving LS paradigm offers a theoretical foundation for undetectable perfect security, realizing it in practice presents several non-trivial challenges, as illustrated in Figure \ref{Fig 2}. In particular, for the same original content at a given embedding position, the stegosystem must construct multiple steganographic coding functions to ensure that any secret message can be embedded by selecting an appropriate coding function. Moreover, during extraction, the stegosystem must reliably identify the coding function used during embedding in order to correctly recover the embedded message.

To implement a practical content-preserving LS system, the following three key challenges must be addressed:
\textbf{1) Construction of Multiple Steganographic Coding Functions:} It is essential to construct more than one steganographic coding function that maps the same original cover content $w_e$ at the embedding position $e$ into different output codes. Formally, $\left| H \right|>1$, and for any two distinct functions $ h, {{h}^{'}}\in H$, it must hold that $h(w_e)\ne {{h}^{'}}(w_e)$. 
This enables the stegosystem to embed different secret messages using the same cover content $w_e$, laying the foundation for content-preserving message embedding.

\textbf{2) Embedding Arbitrary Secret Messages:}
To support the successful embedding of arbitrary secret messages, the stegosystem must be capable of encoding any secret message $m \in M$ by selecting a corresponding steganographic coding function $h \in H$ such that $h(w_e)=m$. This requires that the number of available coding functions satisfies $\left| H \right|\ge {{2}^{n}}$, where 
$n$ is the bit length of the secret message to be embedded. 
For example, to embed all $2^n$ possible $n$-bit messages, there must exist functions $h_0, h_1, \dots, h_{2^n - 1}$ such that $h_i(w_e) = i$.
If this condition is not satisfied, there will exist at least one message $m \in M$ for which no corresponding function $h \in H$ can be found, resulting in embedding failure.

\textbf{3) Correctly Extracting the Secret Message:}
During extraction, the stegosystem must identify the same steganographic coding function $h$ used during embedding in order to decode the embedded secret message correctly. This ensures that the original content $w_e$ at the embedding position $e$ can be used to successfully recover the embedded secret message via $m=h(w_e)$. If a different function $h' \ne h$ is mistakenly applied, the extracted code value $h'(w_e)$ will differ from the original embedded message $m$, resulting in extraction failure.

To address these challenges, we propose a set of solution strategies centered on a unified principle: Prediction Probability Distribution Transformation. This principle leverages the fact that the prediction distributions generated by a Masked Language Model (MLM) can be dynamically adjusted through fine-tuning, enabling flexible steganographic coding without modifying the original text. Distribution transformation forms the foundation for solving the above three challenges under the constraint of content preservation.

\textbf{1) Dynamic Steganographic Coding via Prediction Probability Distribution:}
To construct multiple steganographic coding functions for encoding/decoding the same cover content $w_e$ at a given embedding position, we utilize the prediction distribution produced by an MLM. Specifically, we mask the token at the embedding position within the cover text and use the MLM to predict its replacements, obtaining a probability distribution over the vocabulary. This distribution can serve as the coding domain for the original content. Importantly, the rank of the original content $w_e$ within this distribution determines the message value it can represent. Since this rank varies across different prediction distributions, the same content $w_e$ can be mapped to different secret messages. In this way, different MLM prediction distributions yield different stegnographic coding functions for the same input content.

Because the prediction probability distribution of an MLM is data-driven, fine-tuning the MLM on different datasets alters the distribution. This causes shifts in the rank of the original content word $w_e$ within the distribution, enabling the stegosystem to encode the same content as different secret messages depending on the distribution. These varying distributions thus offer the possibility of constructing multiple steganographic coding functions for the same position. 

\begin{table*}[!htb]
\renewcommand\thetable{5}
    \centering
    \setlength{\tabcolsep}{4pt}
    \begin{tabular}{cm{15cm}c}
    \toprule[1.5pt]
        Method &  \multicolumn{1}{c}{Text} & PPL$\downarrow$ \\ \midrule[1pt]
        Cover & Sometimes, buying in bulk saves us money because it's offered at a cheaper price or we get an item or two for free. & 58.70 \\ \midrule
        FELS & Sometimes, \textbf{selling} in bulk saves us money because it's offered at a \textbf{reasonable} price or we get an \textbf{ounce} or two for free. & 96.70 \\ \midrule
        ARLS & Sometimes, \textbf{selling} in bulk saves us money because it's offered at a \textbf{good} price or we get an item or two for free. & 88.53 \\ \midrule
        CPGLS & Sometimes, buying in bulk saves us money because it's offered at a \textbf{premium} price or we get an item or two for free. & 73.05 \\\midrule
        ADG & As he said an project which is asked there was not one of the most other time we've still says they were interested together. & 475.41 \\\midrule
        Discop & I think when Jeremy sat down and explained why he starts making, he read my Medium Anon site--back then. Only the current story up today, but it's the... & 93.98 \\\midrule[1pt]
        CLstega & Sometimes, buying in bulk saves us money because it's offered at a cheaper price or we get an item or two for free. & \textbf{58.70}\\
    \bottomrule[1.5pt]
    \end{tabular}
    \caption{Generated examples from different linguistic steganography methods}\label{tab3}
\end{table*}

\textbf{2) Controllable Distribution Transformation Through Fine-Tuning the MLM:}
To embed an arbitrary specific secret message, the MLM is fine-tuned in a controlled manner to produce a target prediction distribution, in which the original content $w_e$ appears at a rank corresponding to the desired message value. This controlled distribution transformation ensures that $h(w_e) = m$, where $h$ is the target coding function derived from the fine-tuned distribution and $m$ is the message to be embedded. In other words, by precisely controlling the fine-tuning process, we can construct a prediction distribution that enables the embedding of any arbitrary secret message using the same cover content.

\textbf{3) Accurate Message Extraction Using the Fine-Tuned Target MLM:}
To extract the secret message successfully, we leverage the same fine-tuned MLM alongside the original context to reproduce the prediction distribution at the embedding position. Since the cover text is preserved unchanged during the generation of the stego text, the same fine-tuned MLM generates the identical prediction distribution as in the embedding process. This consistency allows the stegosystem to identify the target steganographic coding function $h$ used for embedding. By applying $h$ to the content $w_e$ at the embedding position, the stegosystem can accurately recover the embedded secret message as $m = h(w_e)$.

Building upon the proposed solution strategies, we introduce a practical linguistic steganographic method, \textbf{CLstega}, which instantiates the content-preserving LS paradigm by leveraging controllable prediction distribution transformation. CLstega supports secret message embedding without modifying the cover text, while ensuring both successful extraction of the hidden secret message and complete consistency between the stego and cover texts.

\subsection{C. Evaluation Metrics}
\textbf{Extraction Success Rate (ESR)} measures the probability of successfully extracting secret messages from stego texts. 
A higher ESR indicates greater reliability and robustness of the extracted secret message, which reflects the feasibility and practicality of the steganographic method. We define ESR as the average ratio of successfully extracted secret message bits to the total embedded bits per sentence:
\begin{equation}\label{eq11}
    \text{ESR}=\frac{1}{N}\sum\limits_{i=1}^{N}{\frac{{{E}_{i}}}{{{B}_{i}}}} ,
\end{equation}
where $N$ represents the total number of sentences, ${{B}_{i}}$ denotes the number of secret message bits embedded in the $i$-th sentence, and ${{E}_{i}}$ denotes the number of bits correctly extracted.

\textbf{Embedding Rate (ER)} is typically used to measure the embedding capacity, which refers to the maximum of secret messages that can be embedded in the cover text while ensuring imperceptibility and security \cite{zhou2021linguistic}. In this paper, ER is defined as the average number of embedded secret message bits per word: 
\begin{equation}
    ER=\frac{1}{N}\underset{i=1}{\overset{N}{\mathop \sum }}\,\frac{{{B}_{i}}}{{{L}_{i}}} ,
\end{equation}
where 
${{L}_{i}}$ represents the number of words in the $i$-th sentence, and ${{B}_{i}}$ denotes the number of secret message bits embedded in the $i$-th sentence.

\textbf{Extraction Time (ET)} measures the computational efficiency of extracting secret messages from the stego texts. A lower extraction time indicates higher extraction efficiency, making the steganographic method more suitable for real-time scenarios. We define ET as the average extraction time per sentence as follows:
\begin{equation}
    ET=\frac{1}{N}\sum\limits_{i=1}^{N}{{{T}_{i}}} ,
\end{equation}
where 
${{T}_{i}}$ denotes the time required to extract the embedded secret messages from the $i$-th sentence.

\textbf{Anti-Steganalysis metrics} evaluate the security of the steganography against detection attacks. Steganalysis techniques often rely on statistical analysis or deep learning models to detect the presence of hidden messages. Following previous work \cite{zhou2021linguistic}, we adopt Accuracy (Acc) and F1 score (F1) of steganalysis classifiers as indicators of security. 

\textbf{Perplexity (PPL)} is a widely used metric to assess the fluency and predictability of text. It quantifies how well a probabilistic language model predicts a given sequence of words, with lower perplexity indicating that the text aligns more naturally with the learned language distribution. In the context of linguistic steganography, perplexity (PPL) 
serves as a measure of imperceptibility, evaluating how natural the stego text appears \cite{zhou2021linguistic}.
A lower PPL value suggests that the stego text maintains high fluency and coherence, making it less distinguishable from normal text and thereby reducing the risk of detection. Conversely, a higher PPL indicates lower text quality and poorer imperceptibility, which could expose the presence of hidden information. PPL is calculated as follows:
\begin{equation}
    PPL={{2}^{-\frac{1}{N}\sum\limits_{i=1}^{N}{\log {{p}_{i}}({{w}_{i1}},{{w}_{i2}},\cdots,{{w}_{in}})}}} ,
\end{equation}
where the word sequence $\{{{w}_{i1}},{{w}_{i2}},\cdots,{{w}_{in}}\}$ denotes the $i$-th sentence, and $p_i(\cdot)$ indicates the probability distribution over the words in the $i$-th sentence as calculated by a language model.

\subsection{D. Case Study}
To further illustrate the perceptual differences among LS methods, Table \ref{tab3} presents generated stego examples along with their Perplexity scores. The cover text remains natural and fluent, while modification-based methods (FELS, ARLS, and CPGLS) introduce slight lexical changes that often alter the original meaning and reduce overall quality. Generation-based methods (ADG and Discop) tend to produce less coherent sentences, resulting in significantly higher PPL values. In contrast, CLstega preserves the exact original cover text, maintaining both the lowest PPL and the highest imperceptibility.

\subsection{E. Limitations}
Our method requires fine-tuning an MLM for embedding, which inevitably incurs additional computational and storage costs compared to classical modification-based approaches. 
It is a trade-off for achieving higher security and avoiding explicit content modification. 
In addition, the current design has not yet explored robustness against textual noise, which will be further investigated in future work.

\end{document}